\documentclass{mn2e}
\input{epsf}

\title[Abell 1689: a complex cluster]
{The~complex~velocity~distribution~of~galaxies~in~Abell~1689: implications for
mass modelling}
\author[E. L. {\L}okas et al.]{E. L. {\L}okas$^{1}$,
F. Prada$^{2}$, R. Wojtak$^{3}$, M. Moles$^{2}$ and S. Gottl\"ober$^{4}$
\\
$^1$Nicolaus Copernicus Astronomical Center, Bartycka 18, 00-716 Warsaw, Poland;
{\rm lokas@camk.edu.pl},\\
$^2$Instituto de Astrof{\'\i}sica de Andalucia (CSIC),
Apartado Correos 3005, E-18080 Granada, Spain \\
$^3$Astronomical Observatory, Jagiellonian University, Orla 171, 30-244 Cracow,
Poland\\
$^4$Astrophysikalisches Institut Potsdam, An der Sternwarte 16, 14482 Potsdam, Germany}

\begin{document}

\maketitle

\vspace{-1in}

\begin{abstract}
The Abell 1689 galaxy cluster has recently become a subject of intensive study. Thanks to its
intermediate redshift ($z=0.183$) its mass distribution can be reconstructed using numerous
methods including gravitational lensing, galaxy kinematics and X-ray imaging.
The methods used to
yield conflicting mass estimates in the past and recently the cluster mass distribution has
been claimed to be in conflict with standard CDM scenarios due to rather large concentration
and steep mass profile
obtained from detailed studies of Broadhurst et al. using lensing. By studying in detail the
kinematics of about 200 galaxies with measured redshifts in the vicinity of the cluster we show
that the cluster is probably surrounded by a few structures, quite distant from each other, but
aligned along the line of sight. We support our arguments by referring to cosmological $N$-body
simulations and showing explicitly that distant, non-interacting
haloes can produce entangled multi-peak line-of-sight velocity distributions similar to that in A1689.
We conclude that it is difficult to estimate the cluster mass reliably
from galaxy kinematics, but the value we obtain after applying a simple cut-off in velocity
agrees roughly with the mass estimated from lensing. The complicated mass distribution around the cluster
may however increase the uncertainty in the determination of the density profile shape obtained with
weak lensing.
\end{abstract}


\vspace{-0.45in}

\section{Introduction}

The increasing amount of data on Abell 1689, a cluster of galaxies at z= 0.183, has recently
motivated several detailed analyses of its dynamical status and mass distribution.
As the largest known gravitational lensing object it has been studied in detail by Broadhurst et
al. (2005a, 2005b) who found that the inferred mass distribution is much steeper compared to
what is expected for dark matter haloes forming in currently available cosmological $N$-body
simulations. In particular they find the concentration parameter of the best-fitting NFW (Navarro,
Frenk \& White 1997) profile of the cluster to be between $c=8$ and $c=14$ depending on projected
radius
to which the mass distribution was studied. The smaller value was obtained from the strong
lensing results in the inner part of the cluster, while the
larger value was found when the study was extended to a scale of 2 Mpc using the
results from weak lensing. The larger value seems rather
high for the estimated mass of $2 \times 10^{15} M_{\sun}$ compared to the expected value of $c=5$
for haloes of this mass as found in cosmological
$N$-body simulations (e.g. Bullock et al. 2001). The discrepancy
has led Oguri et al. (2005) to claim that the mass distribution in A1689 may be in conflict with
the standard cold dark matter (CDM) scenarios for structure formation. They have shown that the
disagreement can be partially reduced if the uncertainties in the parameter estimation due to the
possible triaxiality of the halo are properly taken into account.

The X-ray data for Abell 1689 obtained with the XMM-Newton telescope have been analyzed by
Andersson \& Madejski (2004). The X-ray gas surface brightness distribution appears rather
regular and smooth. However, a closer inspection reveals that the gas temperature profile
is highly asymmetric and the gas mass fraction is lower than usual, which
may point towards a perturbed structure. Moreover, the total
mass inferred from the X-ray analysis gives a value twice as small as
that found from gravitational lensing.

Kinematical analysis of the cluster galaxies shows even larger discrepancies. The redshift survey
performed by Teague, Carter \& Gray (1990) led to the identification of a few structures along the
line of sight and an estimate of 2355 km s$^{-1}$ for the velocity dispersion of the cluster members.
Struble \& Rood (1999) estimate the cluster dispersion to be 1989 km s$^{-1}$ using the same data.
Girardi et al. (1997) applied the wavelet analysis to the same data and detected even more substructure.
Although their estimate for the main cluster velocity dispersion was 1429 km s$^{-1}$, they calculated
the mass by adding the masses of two main substructures with low velocity dispersions of the order of
300-400 km s$^{-1}$ which led to the value of about $2 \times 10^{14} M_{\sun}$, an order of magnitude
lower than the mass obtained from the lensing studies.

In this Letter we reanalyze the velocity distribution of galaxies in the field of A1689
using the larger sample now available. We confirm that the cluster indeed
has a complex structure in velocity space, strongly indicating the presence of
dynamically independent structures along the line of sight. By imposing
different cut-offs in velocity we show
how the cluster mass estimate can change by a large factor, which illustrates
the difficulty in inferring it from the kinematical data.
The complicated mass distribution around the cluster may also affect mass
estimates done with other methods. We refer to cosmological
$N$-body simulations in order to demonstrate that
distant haloes positioned along the line of observations can indeed produce line-of-sight
velocity distributions similar to the one in A1689. Therefore any
estimate of concentration for an object in such environment may be biased by an error not associated
with the method of mass determination but due to the presence of foreground and background structures.

\vspace{-0.25in}

\section{Velocity distribution of galaxies in A1689}

We have searched the NED database for galaxies with redshifts $z=0.1832 \pm 0.05$
and located at distances smaller than 2 Mpc from the cluster centre assumed to be at
RA=$13^{\rm h}11^{\rm m}30.3^{\rm s}$, Dec=$-01^\circ 20'53''$ (J2000). It corresponds to the
position of the elliptical galaxy closest to the centre of the main structure detected
by Girardi et al. (1997).
It is also within 100 kpc from the centre of X-ray gas surface brightness distribution. The
redshift data for galaxies thus chosen come mainly from surveys by Teague et al. (1990),
Balogh et al. (2002) and Duc et al. (2002).

\begin{figure}
\begin{center}
    \leavevmode
    \epsfxsize=8.0cm
    \epsfbox[100 40 480 400]{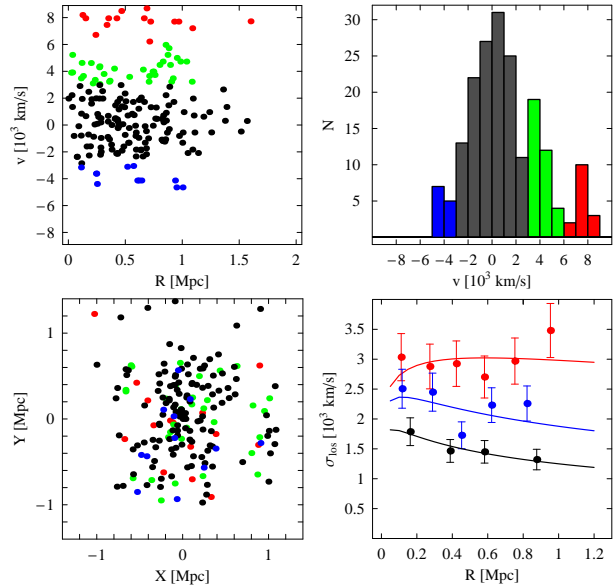}
\end{center}
\caption{Projected distributions of galaxies in the vicinity of A1689. Upper left panel:
line-of-sight velocities of galaxies as a function of projected distance from cluster
centre divided into different
velocity bins $|v| > 6000$ km s$^{-1}$ (red), 3000 km s$^{-1} < |v| < 6000$ km s$^{-1}$ (green and
blue respectively for the positive and negative velocities),
$|v| < 3000$ km s$^{-1}$ (black). Lower left panel: positions of the groups of
galaxies on the surface of
the sky. Upper right panel: the histogram of the line-of-sight velocity distribution
plotting the number
of galaxies per velocity bin of size 1000 km s$^{-1}$ with colour coding as in previous plots.
Lower right panel: line-of-sight velocity dispersion profiles obtained for all galaxies (red),
galaxies with $|v| < 6000$ km s$^{-1}$ (blue) and galaxies with $|v| < 3000$ km s$^{-1}$ (black).
Solid lines show the best-fitting solutions of the Jeans equation.}
\label{a1689vp}
\vspace{-0.25in}
\end{figure}

The line-of-sight
velocities of 192 galaxies in the reference frame of the cluster as a function of distance
from the cluster centre are shown in the upper left panel of Fig.~\ref{a1689vp}. The colours code
the probable membership of the galaxies in different groups separated in velocity space.
The division has been made by a simple cut-off in constant $v$. Separating first galaxies
with $|v| > 6000$ km s$^{-1}$ we get a group of 15 galaxies marked by red dots.
The other two groups with 3000 km s$^{-1} < |v| < 6000$ km s$^{-1}$ are marked with green and
blue respectively for the positive (35 galaxies) and negative velocities (12 galaxies).
The remaining 130 galaxies with
$|v| < 3000$ km s$^{-1}$ are marked with black dots and correspond most probably to the main
body of the cluster. The same colour coding applies to
the velocity distribution histogram (number of galaxies per velocity bin of size 1000 km s$^{-1}$)
shown in the upper right panel of the Figure. The lower left
panel of the Figure shows the position of the galaxies belonging to each group on the plane of
the sky; the positions overlap indicating that the groups lie along the line of sight.

The histogram shown in the upper right panel of Fig.~\ref{a1689vp} is similar to that in Fig.~1
of Girardi et al. (1997), but the identification of structure is somewhat different. In particular, the
green, red and blue peaks in our histogram are the same as those at $c z=60$, $64$ and $52
\times 10^3$ km s$^{-1}$ respectively
in their Figure, but we do not see the structures with velocities close to the cluster mean, which
they identified as S2 and S3, as separate. Although some
fluctuations in the number of galaxies can be seen in this region
when we plot the histogram with better resolution, we
do not think they are significant. The two upper
panels of Fig.~\ref{a1689vp} show qualitatively that the cluster has a complicated structure in
velocity space, at variance with what is expected for relaxed, isolated objects.

The lower right panel of Fig.~\ref{a1689vp} plots the velocity dispersion profiles calculated
using different galaxy samples.
The red profile was obtained from the total sample of 192 galaxies,
for the blue one the galaxies with $|v| > 6000$ km s$^{-1}$ with respect to the cluster
mean were removed, and the black one is
for galaxies with $|v| < 3000$ km s$^{-1}$. The data points were calculated with 30 galaxies
per bin and assigned standard sampling errors (see {\L}okas \& Mamon 2003).
The profiles can be used as a quantitative measure of the
mass of the structure. Assuming that the galaxies trace the overall NFW mass distribution in the
cluster and have isotropic orbits we can estimate the parameters of the NFW profile, the virial
mass $M_v$ and concentration $c$ by fitting the velocity dispersion data to the solutions of
the Jeans equation
\begin{equation}         \label{proj1}
        \sigma_{\rm los}^2 (R) = \frac{2}{I(R)} \int_{R}^{\infty}
	\frac{\nu \sigma_r^2(r) r}{\sqrt{r^2 - R^2}}
	{\rm d} r  \ ,
\end{equation}
where $\nu(r)$ and $I(R)$ are the 3D and the surface distribution of the tracer as a function
of a true ($r$) and projected ($R$) distance from the object centre respectively and
$\sigma_r$ is the radial velocity dispersion related to the mass distribution in the object
(see {\L}okas \& Mamon 2001, 2003).

\begin{table}
\begin{center}
\caption{Best-fitting virial masses and concentrations of A1689
estimated from velocity dispersion profiles
for different galaxy samples. }
\begin{tabular}{cccc}
sample & $M_v [10^{15} M_{\sun}]$ & $r_v$ [Mpc] & $c$    \\
\hline
all galaxies & 33 & 8.3 & 7.3 \\
$|v| < 6000$ km s$^{-1}$ & 7.1 & 5.0 & 22 \\
$|v| < 3000$ km s$^{-1}$ & 2.6 & 3.5 & 28 \\
\hline
\label{parameters}
\end{tabular}
\end{center}
\vspace{-0.25in}
\end{table}

The best-fitting $M_v$ and $c$ values we obtain from the velocity dispersion profiles
for the 3 samples we have considered are given in Table~\ref{parameters}.
The range of mass values illustrates well how much the estimated parameters
depend on the sample of galaxies chosen. We find that
for the whole sample, as well as for the intermediate sample the resulting masses are
significantly larger than expected. Only the most restrictive sample
gives a more reasonable value of $2.6^{+2}_{-1} \times 10^{15} M_{\sun}$ (at 68
per cent confidence level), much more in agreement with the value deduced from recent
studies based on lensing (Broadhurst et al. 2005a,b). Although this sample may still
contain unbound galaxies
which bias the result towards higher masses, it is clear that any further division
of this sample into two parts of comparable size, as was done by Girardi et al (1997),
would result in a mass estimate at least a factor of few lower, strongly at variance with
the value estimated from lensing.

Although our best-fitting concentration for this sample
is much higher than expected, as in the case of studies based on lensing, the data
do not allow us to really constrain the concentration, i.e. all values in the range
$5 < c < 100$ are consistent with the data at $1\sigma$ level. We emphasize,
however, that our cut-offs
in velocity were rather arbitrary and although they followed the gaps in the $v(R)$ diagram
it would be difficult to justify them in a quantitative way.
In particular, none of the galaxies in the $v(R)$
diagram in Fig.~\ref{a1689vp} would be removed by the application of standard methods for the
rejection of outliers. This suggests that in agreement with visual impression from
the $v(R)$ diagram, the galaxies with discrepant velocities are not just interlopers but belong to
some neighbouring structures.

Our analysis illustrates the difficulties encountered when the standard Jeans approach
is uncritically applied to clusters before considering all the possible indications on
their dynamical status and/or their environment. As it happens for A1689, it could be that
discrepant mass estimates are obtained depending on the sample selection criteria.
We note that our sample is a compilation of a few surveys with substantial fraction of
spiral galaxies (41 percent among those with known morphological type). While the sample
of Teague et al. (1990) comes from a standard magnitude-limited survey, the selection
criteria of those of Balogh et al. (2002) and Duc et al. (2002) were aimed at
star-forming galaxies which may bias the sample towards outer regions with more substructure.
The analysis of the cluster dynamics could be significantly improved
if a survey of many galaxy redshifts complete up to a given limiting magnitude was available.
This would allow for a proper comparison
with other well studied clusters and a more accurate description of its velocity distribution.

\vspace{-0.25in}

\section{Comparison with $N$-body simulations}

\begin{table}
\begin{center}
\caption{Best-fitting virial masses and concentrations of the simulated halo
estimated from the 3D information and from velocity dispersion profiles
for different particle samples. }
\begin{tabular}{cccc}
sample & $M_v [10^{14} M_{\sun}]$ & $r_v$ [Mpc] & $c$    \\
\hline
3D information & 5.4 & 2.1 & 9.2 \\
$|v| < 3000$ km s$^{-1}$ & 22 & 3.4 & 5.7 \\
$|v| < 1500$ km s$^{-1}$ & 4.8 & 2.0 & 2.1 \\
\hline
\label{par}
\end{tabular}
\end{center}
\vspace{-0.25in}
\end{table}

In this section we make use of the $N$-body simulations to study the origin of complex velocity
distributions like the one in A1689.
For this work we used the results of a cosmological dark matter simulation described by
Wojtak et al. (2005). The simulation was performed
within a box of size 150 $h^{-1}$ Mpc assuming
the concordance cosmological model ($\Lambda$CDM) with
parameters $\Omega_M=0.3$, $\Omega_{\Lambda}=0.7$, $h=0.7$ and $\sigma_8=0.9$.
The final output of the simulation contained a few tens of massive haloes found with standard
FOF procedures. To mimic observations we place an imaginary observer at a given distance from the
halo and project the particle velocities along the line of sight and their positions
on the surface of the sky.
Among the most massive haloes we chose one with a similar line-of-sight velocity distribution
as is seen in A1689.
The virial mass of the halo is $5.4 \times 10^{14} M_{\sun}$ and it has about $10^4$
particles inside the virial radius which allows us to reliably measure the density profile.
The properties of the halo are summarized in Table~\ref{par}.

\begin{figure}
\begin{center}
    \leavevmode
    \epsfxsize=8.0cm
    \epsfbox[100 30 480 400]{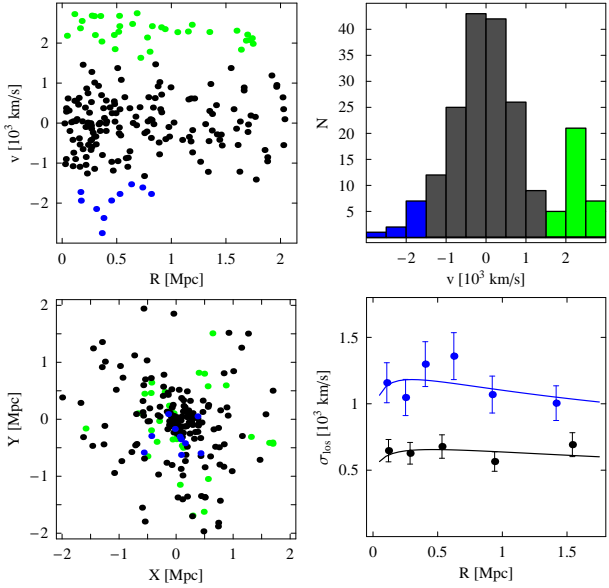}
\end{center}
\caption{Projected distributions of galaxies in the vicinity of the simulated cluster.
Upper left panel:
line-of-sight velocities of 200 halo particles as a function of projected distance from the halo
centre divided into different velocity bins $|v| > 1500$ km s$^{-1}$ (green and
blue respectively for the positive and negative velocities) and
$|v| < 1500$ km s$^{-1}$ (black). Lower left panel: positions of the groups of
particles on the surface of
the sky. Upper right panel: the histogram of the line-of-sight velocity distribution plotting
the number
of particles per velocity bin of size 500 km s$^{-1}$ with colours coded as in previous plots.
Lower right panel: line-of-sight velocity dispersion profiles obtained for all galaxies (blue)
and galaxies with $|v| < 1500$ km s$^{-1}$ (black). Solid lines show the best-fitting
solutions of the Jeans equation.}
\label{halo6vp}
\vspace{-0.2in}
\end{figure}

Out of all particles seen by our observer in the direction of the halo inside the projected
radius $R=r_v$ and with velocities $|v| < 3000$ km s$^{-1}$ with respect to the mean velocity of
the halo we randomly select 200 particles to have a similar statistics as for A1689. (The velocity
range corresponds to about $4 \sigma_{\rm los}$ for a halo of this mass.) The summary
of the observed properties of the halo is presented in Fig.~\ref{halo6vp} with the panels
analogous to those in Fig.~\ref{a1689vp} for A1689. The $v(R)$ diagram in the upper left panel
is highly irregular with particle velocities
spread out over the whole velocity range and making it difficult to decide which of them should be
treated as true members of the halo. A distinct structure is seen at about $v=2500$ km s$^{-1}$
which is even better visible in the upper right panel showing the line-of-sight velocity histogram.
As for A1689 we separate the particles using the velocity criterion; those with
$|v| < 1500$ km s$^{-1}$ are marked with black dots in the plots while those with
$|v| > 1500$ km s$^{-1}$ with green or blue dots depending on the sign of the velocity.
The lower left panel shows the positions of the particles belonging to different velocity bins on
the surface of the sky.

The lower right panel of Fig.~\ref{a1689vp}
plots the velocity dispersion profiles calculated from the data from all
particles (blue) and only from those with $|v| < 1500$ km s$^{-1}$ (black).
As for A1689 we did a similar
exercise of fitting these data with the solutions of the Jeans equation (\ref{proj1}) assuming
isotropic orbits and adjusting the virial mass and concentration. The results for different samples
are listed in Table~\ref{par}. As we can see, the estimated parameters differ dramatically depending
on the sample. For the sample with $|v| < 1500$ km s$^{-1}$ the virial mass is
$M_v = 4.8^{+1.4}_{-1.6} \times 10^{14} M_{\sun}$ (at 68 per cent confidence level) which agrees well
with the mass $M_v = 5.4 \times 10^{14} M_{\sun}$ known from the 3D information about the halo.
The concentration proves more difficult to estimate with such a small sample and simple modelling
since we get
$c=2.1^{+5}_{-1}$ (at 68 per cent c.l.) which does not include the true value $c=9.2$.

\begin{figure}
\begin{center}
    \leavevmode
    \epsfxsize=8.0cm
    \epsfbox[55 55 380 420]{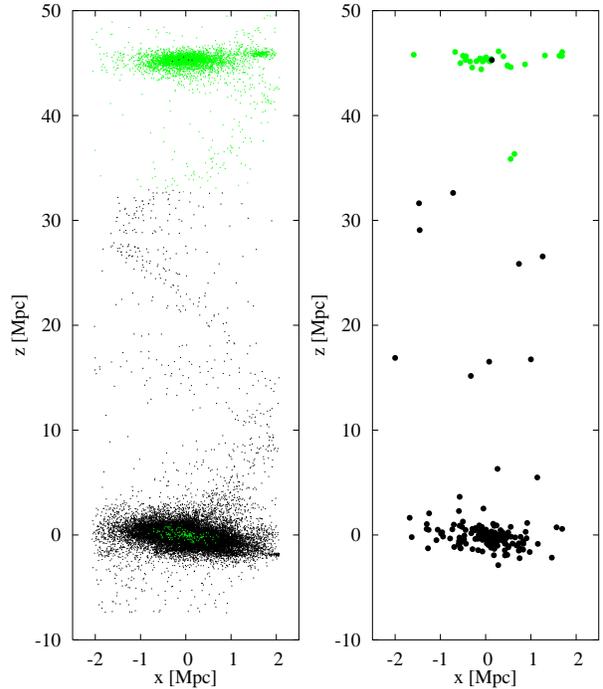}
\end{center}
\caption{Projection of the distribution of dark matter particles in the vicinity of the simulated
halo. The main halo is located at $x=0$, $z=0$. The observation is done along the $z$ axis.
The second, smaller halo is located at the distance of 45 Mpc from the main halo.
The particles with line-of-sight velocities $v < 1500$ km s$^{-1}$ with
respect to the mean velocity of the main halo are coded with black dots, those with larger
velocities with green ones. Left panel shows all particles, while the right one about 200 chosen
randomly to create the mock data. The flattening of the haloes is due to different distance
scales along the two axes.}
\label{map}
\vspace{-0.17in}
\end{figure}

What is the reason behind the complicated velocity structure of the halo? The actual spatial
distribution of the particles in a part of the
observed region is shown in projection in Fig.~\ref{map}.
The centre of our halo is located at $x=0$, $z=0$. The line of sight of the observer is along
the $z$ axis of the plots. The second, smaller halo of mass $M_v = 8 \times 10^{13} M_{\sun}$
contributing to the
$v(R)$ diagram shown in Fig.~\ref{halo6vp} is located at the distance of 45 Mpc from the main halo.
As before, we marked the particles with line-of-sight velocities $v < 1500$ km s$^{-1}$ (with
respect to the mean velocity of the main halo) with black dots and those with larger
velocities with green ones. Left panel shows all particles in this region of the simulation box,
while the right one about 200 chosen randomly to create the mock data (not all 200 particles are
shown because the region in the direction of negative $z$ is not plotted).

The Figure demonstrates that in spite of their proximity in velocity space ($2300 $ km s$^{-1}$
is of the order of $3 \sigma_{\rm los}$ of the big halo) the two haloes are in fact very distant.
The distance of the smaller halo, 45 Mpc, corresponds to about 20 virial radii of the big halo
therefore the haloes do not affect each other dynamically and are not bound to each other, but
still their projected velocity distributions are entangled (note that there are green particles
in the centre of the bigger halo and black particles in the centre of the smaller halo).
The reason for this is the rather low value of
the Hubble velocity in comparison with velocity dispersion of bound structures, e.g.
velocity dispersions of massive haloes or galaxy clusters are comparable to the Hubble flow at
distances as large as 8 virial radii from their centres (see Fig. 3 of Wojtak et al. 2005).
The smaller halo in our example is receding from the bigger one with a velocity of about
$2300 $ km s$^{-1}$ mainly due to the Hubble flow which at distance of 45 Mpc is
of the order of $3000 $ km s$^{-1}$.

\begin{figure}
\begin{center}
    \leavevmode
    \epsfxsize=8.0cm
    \epsfbox[100 30 480 200]{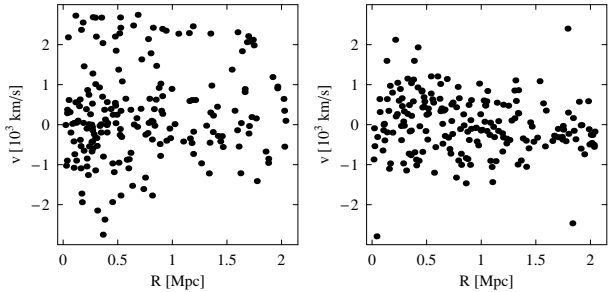}
\end{center}
\caption{Comparison of line-of-sight velocities of dark matter particles as a function of projected
distance from halo centre. Left panel shows the same diagram as in upper left panel of
Fig.~\ref{halo6vp}.
Right panel shows the $v(R)$ diagram for the same halo observed in a different direction.
Both panels plot 200 randomly selected particles.}
\label{comparh6}
\vspace{-0.2in}
\end{figure}

To further illustrate the point, we provide an example of a well-behaved $v(R)$
diagram. Fig.~\ref{comparh6} shows again the $v(R)$ diagram of our simulated halo
from Fig.~\ref{halo6vp} in the left panel, while in the right panel we present an analogous plot of
line-of-sight velocities as a function of projected distance for 200 particles chosen from
the same halo, but observed in a
different direction. In this case the halo has no massive neighbours along the line of sight
and the single particles with
discrepant velocities can be easily dealt with using standard procedures for interloper removal.

Our purposely chosen example illustrates well the difficulties in interpreting the measured
line-of-sight velocities of galaxy clusters in the case of presence of neighbouring structures.
Whereas velocity differences amounting up to $ 2300 $ km s$^{-1}$, as in our simulated haloes,
could easily be interpreted as orbital velocities of dynamically bound objects, they could also
correspond to structures seen in projection but otherwise separated by distances
much larger than their virial radii and, therefore, totally unrelated.

To make a connection with the studies based on lensing we note that the surface density distribution
(the main lensing observable) measured along the line of sight in our simulations is increased
by about 25 percent everywhere along the projected radius of the bigger halo
due to the presence of the smaller halo. The significance of this effect will of course depend on
the exact properties of the haloes and probability of their alignment, which will be studied
elsewhere.

\vspace{-0.25in}

\section{Conclusions}

We have shown that cosmological structures quite distant from each other, when aligned
with the direction of observation, can produce projected velocity distributions which are quite
difficult to interpret. In particular, such extended distributions can lead to very different
velocity dispersion, and therefore mass, estimates. The complicated structure of the velocity
distribution, with many peaks, suggests however that we indeed deal with multiple objects
situated along the line of sight. On the other hand the simulations show that close neighbours,
within one virial radius from each other (like mergers or infalling subhalo), have similar
velocities and would produce regular, one-peak velocity distributions.

Given the multi-peak velocity structure of A1689 we conclude that the groups of galaxies with
$\pm 4000$ km s$^{-1}$ (and of course also the more discrepant group at $+8000$ km s$^{-1}$)
with respect to the cluster mean velocity are probably separate structures not associated with
the cluster, but aligned along the line of sight. If the velocities
$\pm 4000$ km s$^{-1}$ are due mainly to the Hubble flow these structures are located at
about 60 Mpc from the cluster. For a cluster mass of $2 \times 10^{15} M_{\sun}$ the distance
corresponds to about 17 virial radii. This would mean that the structures do not affect the
cluster dynamically and cannot be responsible for any departures from equilibrium.
This rather complex structure in velocity does not necessarily translate itself into the
X-ray gas distribution that can appear regular and smooth. This is indeed the case for
A1689, in which the morphology of the X-ray data is commonly interpreted as a clear indication
of the relaxed state of the cluster. Andersson \& Madejski (2004) demonstrate however that in the
case of two similar clusters aligned along the line of sight X-ray data can easily underestimate
the mass by a factor of 2.

The presence of foreground and background structures in the line of
sight of A1689 may affect the path of light coming from the lensed galaxies.
In their study of strong lensing in A1689 Broadhurst et al. (2005a) managed to
subtract the neighbouring structure from the main lensing signal. However, no such
correction was made when the analysis was extended by Broadhurst et al. (2005b)
to weak lensing and larger distances from the cluster centre.
Hoekstra (2003) has shown that even the presence of distant large-scale structure
in the Universe can affect the weak lensing signal, significantly increasing
the uncertainty in the estimated parameters for a given cluster. A similar increase in
the estimated errors was shown to be the case if the cluster departs from
spherical symmetry (Oguri et al. 2005). It would be interesting to verify whether
objects in the vicinity of
the cluster could cause similar effect, thereby decreasing the claimed discrepancy with
CDM structures, especially when the weak lensing
signal is very low and the inferred surface mass distribution in the outer regions
very uncertain, as in the case of A1689 (Broadhurst et al. 2005b).

\vspace{-0.25in}

\section*{Acknowledgements}

We wish to thank T. Broadhurst, H. Hoekstra and the anonymous referee for their comments
on the paper.
Computer simulations presented in this paper were performed at the
Leibnizrechenzentrum (LRZ) in Munich.
E{\L} is grateful for the hospitality of Instituto de Astrof{\'\i}sica de Andalucia
in Granada where part of this work was done.
RW acknowledges the summer student program at Copernicus Center.
This research has made use of the NASA/IPAC Extragalactic Database (NED)
operated by the Jet Propulsion Laboratory.
This work was partially supported by the
Polish Ministry of Scientific Research and Information Technology
under grant 1P03D02726 and the exchange program of CSIC/PAN.

\end{document}